\documentclass[
twocolumn,
prd,
 preprintnumbers,
superscriptaddress,
 amsmath,amssymb,
 aps,
]{revtex4-1}
\pdfoutput=1
\usepackage[colorlinks,linktocpage,hypertexnames=false]{hyperref}
\hypersetup{colorlinks=true, citecolor=blue}
\usepackage{graphicx}

\usepackage[utf8]{inputenc}
\usepackage[dvipsnames]{xcolor}
\usepackage{physics}
\usepackage{natbib}

\newcommand{\be}{\begin{equation}}
\newcommand{\ee}{\end{equation}}
\newcommand{\ba}{\begin{eqnarray}}
\newcommand{\ea}{\end{eqnarray}}

\begin{document}

\preprint{INR-TH-2026-008}

\title{False vacuum decay around black holes: \\ calculation from first principles in the (3+1)-dimensional case}

\author{Ratmir Gazizov}
\email{gazizov.rl18@physics.msu.ru}
\affiliation{Institute for Nuclear Research of the Russian Academy of Sciences, 117312 Moscow, Russia}
\affiliation{Physics Faculty, Lomonosov Moscow State University, 119991 Moscow, Russia}

\author{Dmitry Gorbunov}
\email{gorby@ms2.inr.ac.ru}
\affiliation{Institute for Nuclear Research of the Russian Academy of Sciences, 117312 Moscow, Russia}
\affiliation{Moscow Institute of Physics and Technology, 141700 Dolgoprudny, Russia}

\author{Dmitry Levkov}
\thanks{Deceased.}
\affiliation{Institute for Nuclear Research of the Russian Academy of Sciences, 117312 Moscow, Russia}
\affiliation{Institute for Theoretical and Mathematical 
Physics, MSU, Moscow 119991, Russia}

\date{August 10, 2026}

\begin{abstract}
In $(3+1)$ spacetime dimensions, we numerically calculate the suppression exponent of the false vacuum decay probability in the presence of a thermal bath and a black hole. 
\end{abstract}

\maketitle


{\it 1.} According to experimental data and calculations within the Standard Model, the electroweak vacuum is metastable \cite{Buttazzo_2013} and can decay through quantum tunneling. In flat space, the decay probability is exponentially suppressed, and the lifetime of the electroweak vacuum much exceeds the age of the Universe \cite{Andreassen:2017rzq}.

However, it has been suggested that sufficiently small black holes (BHs), acting as impurities, can significantly increase the decay probability in their vicinity because of their high temperature $T_{\text{BH}}$ \cite{BEREZIN1988397, Hiscock:1987hn}. If this argument is correct, then the corresponding conjecture about the decay probability must be valid for any model with a false vacuum potential. Would  BHs be in thermal equilibrium, the exponential suppression may disappear \cite{Burda_2016}. However, in a realistic case BHs do not equilibrate.

In the consistent approach from first principles, the stationary initial state includes a BH and outgoing Hawking radiation. The decay probability is derived from the path integral in the saddle-point approximation. The probability is then exponentially suppressed,  
\begin{equation}
\label{semi-exp}
\mathcal{P} \sim e^{-F}.
\end{equation}
with a specific functional of the saddle solution $F$. The same method was applied to various induced processes; see e.g. \cite{Khlebnikov:1991th, Kuznetsov_1997}.

We consider a model with simple scalar potential:
\begin{equation}\label{eq:potential}
	V(\phi) = \frac{1}{2} m^2 \phi^2 -  \frac{1}{2} g m \phi^3\,.
\end{equation}
Similarly to the SM Higgs sector, this model has a false vacuum at small field values and a true vacuum at large values of the field~$\phi$. 
We obtain numerically the  
semiclassical solutions for this model to calculate the decay probability around BHs of various masses. The decay probability remains exponentially suppressed for all, even small, BH masses. We explain the physics behind this phenomenon, that differs from what was suggested on example of (1+1)-dimensional toy models \cite{Shkerin_2021}.  


{\it 2.} We consider a scalar field $\phi$ in a static Schwarzschild spacetime with the action
\begin{equation}\label{eq:actsch}
	S = \int d^4x \sqrt{-\det g_{\mu\nu}} 
	\left[ -
	\frac{1}{2} \phi\, g_{\mu\nu}\nabla^\mu\nabla^\nu   \phi - V(\phi) 
	\right],
\end{equation}
where $g_{\mu\nu}$ is the external Schwarzschild metric,
\begin{equation}\label{eq:Sch}
	ds^2 \!=\! f(r)dt^2 \!-\! \frac{1}{f(r)}dr^2 \!-\! r^2 d\Omega^2\,, 
	\qquad f(r) =1-\frac{r_h}{r}\,.
\end{equation}
Here, $r_h$ denotes the horizon radius. We neglect any backreaction of the field, which energy is much lower than the BH mass, $E_{\mathrm{field}} \ll M_{\mathrm{BH}}$. Indeed, our numerically obtained configurations exhibit $E_{\mathrm{field}} \sim m/g^2 \ll r_h M_{Pl}^2/2$. 

The initial field state of an evaporating BH in empty space is the Unruh vacuum, described by the density matrix $\hat{\rho}_R \sim e^{-\hat{H}_R / T_{\mathrm{BH}} }$, where $T_{\mathrm{BH}}$ and $\hat{H}_R$ are the temperature and Hamiltonian of the outgoing modes~\cite{jacobson2004introductionquantumfieldscurved}. Similarly, the state of incoming flux at temperature $T_{\infty}$ is described by the density matrix $\hat{\rho}_L \sim e^{-\hat{H}_L / T_{\infty}}$. The total initial state is $\hat{\rho} = \hat{\rho}_R \otimes \hat{\rho}_L$, and 
the final state after vacuum decay $|\phi_{\mathrm{f}} \rangle$ is any field configuration  $\phi_{\mathrm{f}}(\mathbf{x})$ with an expanding true-vacuum bubble.

The decay probability is obtained via the path integral
\begin{equation}\label{eq:probpath}
	\mathcal{P} \sim \int_{\phi_{\mathrm{f}} \in \mathrm{TV}} \mathcal{D}\phi_{\mathrm{f}}\,
	\langle \phi_{\mathrm{f}} | \hat{\mathcal{S}}\hat{\rho}\hat{\mathcal{S}}^\dagger | \phi_{\mathrm{f}} \rangle\,,
\end{equation}
with $\hat{\mathcal{S}}$-matrix. The integration is taken over all configurations $\phi_{\mathrm{f}}$ in the basin of attraction of the true vacuum.

Assuming that the interesting configurations are spherically symmetric, we introduce new variables:
\begin{equation}
\label{eq:varphi}
	\varphi \equiv g\phi r\,, \,\, dr/dx \equiv f(r)\,, \,\,  x \equiv r + r_h \log \left( r/r_h - 1\right).
\end{equation}
Then action \eqref{eq:actionV} takes the form:
\begin{align}\label{eq:actionV}
	&S \!=\! \frac{4\pi}{g^2}\!\!\! \int\!\!\! dt dx\!\!
	\left[ -
	\frac{\varphi}{2}  (\partial_t^2 \varphi \!-\! \partial_x^2 \varphi) \!-\!
	\frac{\varphi2}{2} U_{\mathrm{eff}}(x) \!+\! \frac{f(r)}{2r}m\varphi^3
	\right],
\\\nonumber
	&U_{\mathrm{eff}}(x) = \frac{f(r)f'(r)}{r} + m^2 f(r)\,.
\end{align}

There are three independent dimensionless parameters ($r_hm$, $T_{\mathrm{BH}}/m$, $T_{\infty}/m$), the action \eqref{eq:actionV}  depends on $r_h$, and the initial state contains right- and left-moving thermal fluxes at different temperatures $T_{\mathrm{BH}}$ and $T_{\infty}$. If $T_{\mathrm{BH}} = 1/4\pi r_h$, then the problem describes an evaporating BH in a thermal bath at temperature $T_{\infty}$. In particular, if $T_{\infty}=0$, it describes the Unruh vacuum, while if $T_{\infty}=T_{\mathrm{BH}}$, it describes the Hartle-Hawking vacuum.

\begin{figure}[h]
	\centering
    \includegraphics[width=\columnwidth]{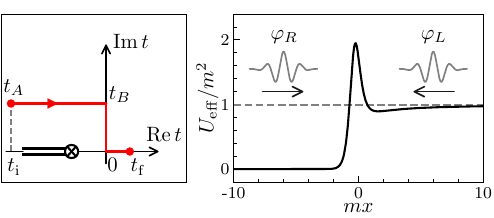}
    \caption{Left: contour $\{t_\mathrm{i} t_A t_B 0 t_\mathrm{f}\}$ in the complex time plane. The thick black line and cross denote the branch cut and branch point of the analytic continuation of the solution $\varphi_{\mathrm{cl}}(t,x)$. Right: black and gray lines show function $U_{\mathrm{eff}}(x)$ for $mr_h = 0.15$ and schematic right- and left-moving wave packets $\varphi_R$ and $\varphi_L$.}
	\label{fig:contour}
\end{figure}

{\it 3.} The decay probability \eqref{eq:probpath} can be estimated in the saddle-point approximation\,\eqref{semi-exp}, if $g \ll 1$, see \cite{Shkerin_2021} for the derivation. In our case the path integral\,\eqref{eq:probpath} is saturated by the saddle solution $\varphi_{\mathrm{cl}}(t,x)$ to the field equation
\begin{equation}\label{eq:fieldeq}
	\partial^2_t \varphi_{\mathrm{cl}} - \partial^2_x \varphi_{\mathrm{cl}} + U_{\mathrm{eff}}(x)\varphi_{\mathrm{cl}}-\frac{3f(r)}{2r}m\varphi_{\mathrm{cl}}^2=0\,.
\end{equation}
It is defined on the complex contour $\{t_\mathrm{i} t_A t_B 0 t_\mathrm{f}\}$ shown in Fig.~\ref{fig:contour}, where $|t_{\mathrm{f}} - t_{\mathrm{i}}| \rightarrow \infty$, and hence is complex-valued.

At $t=t_{\mathrm{i}} \rightarrow - \infty$, the field $\varphi_{\mathrm{cl}}$ approaches the false vacuum, and all perturbations above the false vacuum are located in the regions $x \rightarrow \pm \infty$, where the equation is linear,
\begin{equation}\label{eq:lineq}
	\partial^2_t \varphi_{\mathrm{cl}} - \partial^2_x \varphi_{\mathrm{cl}} + U_{\mathrm{eff}}(x)\varphi_{\mathrm{cl}}=0\,.
\end{equation}
Its solution is a sum of right- and left-moving modes, 
\begin{equation}
\label{eq:fielddecomp}
\begin{split}
    \varphi_{\mathrm{cl}}(t_{\mathrm{i}}, x)=\varphi_{R}(t_{\mathrm{i}}, x) + \varphi_{L}(t_{\mathrm{i}}, x)  \\
	=\!\!\!\!\sum_{\! I=R,L} \!\!\int\!\!\! \frac{d\omega}{\sqrt{4\pi\omega}} 
	 \!\left[
	a_{I,\omega}f_{I,\omega}(x)e^{-i\omega t_{\mathrm{i}}}\!+\!b_{I,\omega}f^*_{I,\omega}(x)e^{i\omega t_{\mathrm{i}}}
	\right]\!\!,
    \end{split}
\end{equation}
with positive(negative)-frequency amplitudes $a_{I,\omega}$($b_{I,\omega}$), see Fig.~\ref{fig:contour}. They are semiclassical analogues of annihilation and creation operators. Functions $f_{I,\omega}(x)$ are right- and left-moving modes obeying the equation obtained from (\ref{eq:lineq}) upon substitution $\varphi_{\mathrm{cl}}=e^{- i\omega t}f_{I,\omega}(x)$:
\begin{equation}\label{eq:modeeq}
	-f_{I,\omega}''(x)+U_{\mathrm{eff}}(x)f_{I,\omega}(x)=\omega^2f_{I,\omega}(x)\,,
\end{equation}
which is analogous to a stationary Schr\"odinger equation. These orthogonal and normalized to unity modes evolve as $f_{L,\omega}\!\sim\!e^{-i \omega x}$ at 
$x\to -\infty$ and $f_{R,\omega}\!\sim\! e^{ikx}$ at $x\to +\infty$, where $k=\sqrt{\omega^2 - m^2}$. There are only right-moving modes for $0 < \omega \leq m$.

The initial conditions imposed on the amplitudes are
\begin{equation}\label{eq:BCinitial}
		a_{L,\omega} = e^{-\beta_{\infty} \omega} (b_{L,\omega})^*, \;\;\;\;\;\;
		a_{R,\omega} = e^{-\beta_{\mathrm{BH}} \omega} (b_{R,\omega})^*,
\end{equation}
where $\beta_{\infty}\equiv 1/T_\infty$ and $\beta_{\mathrm{BH}}\equiv 1/T_\mathrm{BH}$. At $t_{\mathrm{f}} \rightarrow +\infty$ the field $\varphi_{\mathrm{cl}}$ must belong to the basin of attraction of the true vacuum. We impose the following final conditions:
\begin{equation}\label{eq:BCfin}
	\mathrm{Im}\,\varphi_{\mathrm{cl}}(t_{\mathrm{f}},x)=\mathrm{Im}\,\partial_t\varphi_{\mathrm{cl}}(t_{\mathrm{f}},x) =0\,.
\end{equation}
They correspond to the inclusive final state in (\ref{eq:probpath}).

Since eq.\,(\ref{eq:fieldeq}) is linear in contour segment $\{t_{\mathrm{i}}t_A\}$ when $t_{\mathrm{i}} \rightarrow -\infty$, and solution $\varphi_{\mathrm{cl}}(t,x)$ is analytic in the complex time plane, we can concentrate on the contour ${\cal C}\equiv\{t_A t_B 0 t_{\mathrm{f}}\}$ in Fig.~\ref{fig:contour}. 
The suppression $F$ in \eqref{semi-exp} becomes a functional \cite{Shkerin_2021} of semiclassical solution $\varphi_{\mathrm{cl}}(t,x)$:
\begin{equation}
\begin{split}
\label{eq:suppint}
&F=2\mathrm{Im}\,S 
    \\ 
	=& 2\mathrm{Im}\!\left[ \frac{4\pi}{g^2}\!\!\!\int \!\!\!dt\!\!\!\int^{\infty}_{-\infty}\!\!\!\!\!\!\!\!\!dx\,
	\!r^2\!f(r)
	\!\left(\!
	\frac{1}{2}\varphi_{\mathrm{cl}}V_{\mathrm{int}}'\!\left(\!\!\frac{\varphi_{\mathrm{cl}}}{r}\!\right) \!-\! V_{\mathrm{int}}\!\left(\!\frac{\varphi_{\mathrm{cl}}}{r}\!\right)\!\! 
	\right)\!\right]
\end{split}
\end{equation}
where the time integral is taken along the contour $\mathcal{C}$, and $V_{\mathrm{int}}(\phi)$ is the cubic part of potential (\ref{eq:potential}).


{\it 4.} We solve the boundary-value problem numerically 
and adopt units with $m=1$. We introduce an $N_t \times N_x$ lattice $\{t_i, x_j \}$ that covers the domain $-L \leq x \leq L$ and the contour $\mathcal{C}$ with finite $t_{\mathrm{i}}$ and $t_{\mathrm{f}}$. The $x$ lattice has a uniform step $\Delta x$, while the $t$ lattice has a uniform step $\Delta t$ separately on the intervals $\{t_A t_B\}$, $\{t_B 0\}$, and $\{0 t_{\mathrm{f}}\}$. The field is defined on lattice sites $\varphi_{i,j}=\varphi(t_i,x_j)$. Typical parameter values are $L = 50\ldots 100$, $(t_{\mathrm{f}} - t_{\mathrm{i}}) = 25\ldots 30$, $\Delta x = 0.05\ldots 0.1$, and $\Delta t = 0.025\ldots 0.05$. We use the standard second-order discretization of the problem and arrive at the system of nonlinear algebraic equations for variables $\varphi_{i,j}$.

The system is solved with the Newton--Raphson method: given an initial guess $\varphi^{(0)}_{i,j}$, we substitute $\varphi_{i,j} = \varphi^{(0)}_{i,j} + u^{(0)}_{i,j}$ into the equations and the boundary conditions. It provides a system of $N_t\times(N_x-2)$ complex linearized equations for the variables $u^{(0)}_{i,j}$, which form a sparse linear system. To solve it, we use the methods described in \cite{Bonini_1999, Demidov_2015}. Once the system is solved for $u^{(0)}$, we set our new guess, $\varphi^{(1)}_{i,j} = \varphi^{(0)}_{i,j} + u^{(0)}_{i,j}$, and iterate until the configuration $\varphi^{(n)}_{i,j}$ converges to the required accuracy.

The Newton-Raphson method also allows us to get solutions for different values of the model parameters, since their small changes are not expected to noticeably change the solution. We use the solution for parameters $\beta_{\mathrm{BH}}$, $\beta_{\infty}$, and $r_h$ as a seed to find the solution for slightly  different $\beta_{\mathrm{BH}} + \Delta \beta_{\mathrm{BH}}$, $\beta_{\infty} + \Delta \beta_{\infty}$, and $r_h + \Delta r_h$.

The numerical errors associated with the lattice are estimated performing extrapolation $\Delta x,\, \Delta t,\, 1/L\rightarrow 0$. The typical relative numerical errors $\delta F / F$ are $\delta F / F < 0.3\%$ for $\Delta x$, $\delta F / F < 0.1\%$ for $\Delta t$, and $\delta F / F < 0.2\%$ for $L$. Energy is conserved with accuracy $\delta E / E < 2 \%$. The linearization at $t = t_{\mathrm{i}}$ is accurate to $ |E_{\mathrm{lin}} - E| / E < 0.3\%$, where $E_{\mathrm{lin}}$ is the energy calculated with only the quadratic part of the potential.


{\it 5.} We start with the simple case $\beta_{\mathrm{BH}} = \beta_{\infty}$, when analyticity of the solution (\ref{eq:fielddecomp}) on the interval $\{t_{\mathrm{i}} t_A\}$ and initial conditions (\ref{eq:BCinitial}) imply that the field is real at initial time $t_A = t_i + i\beta_{\infty} /2$. Due to boundary conditions (\ref{eq:BCinitial}), (\ref{eq:BCfin}) and equation (\ref{eq:fieldeq}), the solution is real on  intervals $\{t_A t_B\}$ and $\{0t_{\mathrm{f}}\}$. Let us assume that $ \mathrm{Re}\, \partial_t \varphi = 0$ at $t_B$ and $0$. Then the solution is also real on interval $\{t_B 0\}$ and has turning points $\partial_t \varphi = 0$ at $t_B$ and $0$. This is a periodic solution in Euclidean time. Since only the imaginary part of the action contributes to the suppression (\ref{eq:suppint}), only the Euclidean part of the solution needs to be considered; this solution is called a periodic instanton, and its period equals the inverse temperature.

\begin{figure}[!htb]
	\centering
    \includegraphics[width=\columnwidth]{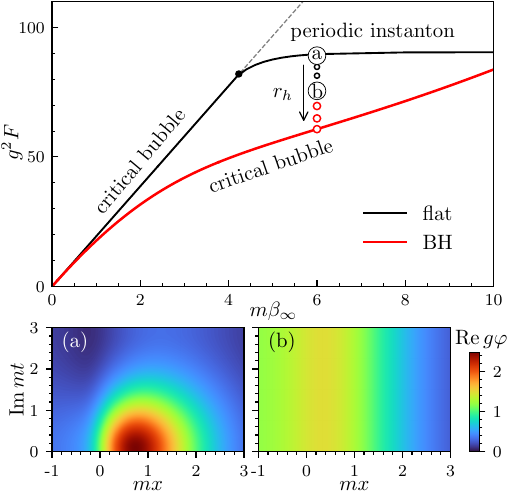}
	\caption{Top: decay suppression as a function of $\beta_\infty$ for the vacuum in Minkowski spacetime (black line) and the Hartle-Hawking vacuum (red line). The gray dashed line shows the critical-bubble branch with decay suppression $F= \beta_{\infty} E_{\mathrm{crit}}$. The circles show suppressions of periodic instantons calculated for fixed $m\beta_{\infty}=6$ and various $r_h$. Black circles correspond to non-trivial periodic instantons; red circles show static solutions, i.e., critical bubbles. Bottom: examples of the solutions ($N_t \times N_x = 31 \times 601$, $mL=30$). (a) A periodic instanton for $mr_h = 0.1$. (b) A periodic instanton that has degenerated into the critical bubble for $mr_h = 0.32$.}
	\label{fig:flat-HH}
\end{figure}

We begin by calculating periodic instantons in the flat spacetime to test the numerical method. We consider periodic instantons in the period range $m \beta_{\infty} = 0\ldots 10$. The suppression as a function of inverse temperature is shown by the black line in Fig.~\ref{fig:flat-HH}. The instanton at large period $\beta_{\infty}$ can be calculated with the shooting method, and instantons for other periods can be obtained by step-by-step small changes of the period $\beta_{\infty}$. At sufficiently high temperature, the decay is described by the static solution, which is the critical bubble. 
It corresponds to thermal vacuum decay with probability $\mathcal{P} \sim e^{-\beta_{\infty} E_{\mathrm{crit}}}$. The opposite limit of large period corresponds to quantum tunneling at zero temperature, with probability $\mathcal{P} \sim e^{-F_0}$, where $g^2 F_0 \simeq 91$.

Further, we adopt periodic instanton $\varphi_{\mathrm{inst}}(t,r)$ with period $m\beta_{\infty}$ as an approximate solution $\varphi_{\mathrm{inst}}(t,r(x))$ in the generalized problem with initial-condition parameters $\beta_{\mathrm{BH}} = \beta_{\infty}$ and potential parameter $mr_h \ll 1$, and apply the Newton-Raphson method to find the exact solution for these parameters (Fig.~\ref{fig:flat-HH}a). 
Then, the potential can be changed in small steps until $r_h = 1/4 \pi T_{\infty}$ is reached and the solution describing the decay of the Hartle-Hawking vacuum is found. We observed that the solution degenerates into the critical bubble as $r_h$ grows (Fig.~\ref{fig:flat-HH}b), in agreement with known results \cite{Gregory_2014, Shkerin_2021}. 
In our model, the thermal vacuum decay over the entire range of considered BH sizes proceeds via the critical bubble. Its probability is $\mathcal{P} \sim \exp(-\beta_{\infty} E_{\mathrm{crit}}(r_h))$, see the red line in Fig.~\ref{fig:flat-HH} for the suppression function.

Therefore, in the range $m\beta_{\infty} = 0 \ldots 10$, the BHs located in a thermal bath of the same temperature act as catalysts, since the suppression $F$ decreases compared to the case without a BH. However, for small BHs with $m\beta_{\mathrm{BH}} \leq 1$, the two suppression functions converge. This happens because small BHs do not significantly change the geometry of spacetime, so the corresponding solutions also coincide. In this limit, the decay is driven mainly by the high temperature of the environment. 


{\it 6.} However, realistic BHs are not in thermal equilibrium with their environment. We therefore need to understand how the result changes when the environment temperature is set to zero. The region of small BHs is of particular interest: in this limit, we ask whether the decay probability remains exponentially suppressed.

\begin{figure}[!htb]
	\centering
    \includegraphics[width=\columnwidth]{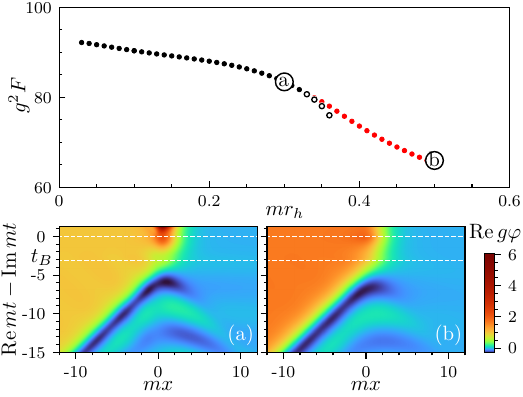}
    \caption{Top: decay suppression dependence on $r_h$ at $m\beta_{\infty} = 10^6$ and  $m\beta_{\mathrm{BH}}/4\pi = 0.5$. Black dots are solutions obtained without $\epsilon$-regularization, but the algorithm starts to diverge, as shown by black circles. Red dots are solutions with $\epsilon$-regularization. Bottom: (a) The solution for $mr_h = 0.3$ ($N_t \times N_x = 476\times 601$, $mL = 30 $). (b) The solution for $mr_h = m\beta_{\mathrm{BH}}/4\pi = 0.5$ with critical-bubble creation ($N_t \times N_x = 609 \times 2001$, $mL = 100 $). It describes the decay of  Unruh vacuum around a BH with horizon radius $r_h$.}
	\label{fig:mink-U}
\end{figure}

The recipe of the study above can be used to find solutions of the generalized problem in the case $T_{\infty}/m = 0$, but then the full contour in the complex time plane must be considered. Therefore, the periodic instanton $\varphi_{\mathrm{inst}}(t,r(x))$ for $\beta_{\mathrm{BH}} = \beta_{\infty}$ and $mr_h \ll 1$ must be continued to the complex time contour using the equation and then adopted as an approximate solution for the Newton-Raphson method. After that, we rise $m\beta_{\infty}$ from $m\beta_{\mathrm{BH}}$ to $10^6$. The potential can then be changed in small steps to set $r_h = \beta_{\mathrm{BH}}/(4 \pi)$, as shown in Fig.~\ref{fig:mink-U} with solutions of the form in Fig.~\ref{fig:mink-U}a.

As $r_h$ grows, the final true-vacuum bubble in the solutions tends to the critical bubble. Unfortunately, this causes our algorithm to fail (black circles in Fig.~\ref{fig:mink-U}), since the critical bubble is a static solution and tends to occupy the entire finite lattice. Analytically, the critical bubble should be created in the infinite future, $t_{\mathrm{f}} \rightarrow \infty$.

To solve this problem, we use the $\epsilon$-regularization described and justified in \cite{Bezrukov_2004, Demidov_2015}. We add a small imaginary term to the action:
\begin{align}
	S &\rightarrow S_{\epsilon} = S[\varphi] + i\epsilon T_{\mathrm{int}}[\varphi]\,,  \qquad 0 < \epsilon \ll 1\\
\label{eq:Tint}
	T_{\mathrm{int}}[\varphi] &= \frac{1}{g^2} \int dt dx\, G(x) W_{\mathrm{int}}(\varphi / a \varphi_{\mathrm{max}})\,,
\end{align}
where $G(x)=e^{-(x-x_{\mathrm{max}})^2/2\sigma^2}$,  $W_{\mathrm{int}}(u) = u^4 e^{-u^2}$, $\varphi_{\mathrm{max}}$~is the maximum field value of the critical bubble, and its coordinate after decay at $t=t_{\mathrm{f}}$ is $x_{\mathrm{max}}$. We set $\sigma = 0.4$ and $a=0.5$ in the numerical calculations. The function $G(x)$ restricts the spatial integral to the region where the field starts to grow rapidly after tunneling. The function $W_{\mathrm{int}}$ has positive values in $0 < \varphi < \varphi_{\mathrm{max}}$ and is vanishingly small at $\varphi \simeq 0$ and $\varphi > \varphi_{\mathrm{max}}$. The time integral in (\ref{eq:Tint}) diverges if the solution contains a static critical bubble at $t\rightarrow \infty$, and such a solution cannot be an extreme of the regularized action $S_{\epsilon}$. However, this time integral converges if the solution contains an expanding bubble, and the regularized action can have extrema at such configurations.

The $\epsilon$-regularization makes the critical bubble non-static because of the slightly modified equations: it decays to the true vacuum in a finite time. As $\epsilon \rightarrow 0$, the lifetime of the critical bubble tends to infinity, and the solutions continuously approach the correct solution at $\epsilon = 0$. Moreover, the suppression of a solution with $\epsilon$-regularization can be written as $F_{\epsilon}=F_{\epsilon=0}+\mathcal{O}(\epsilon)$. We perform a linear extrapolation in the limit $\epsilon \rightarrow 0$ to obtain almost accurate values of the suppression without regularization, $F_{\epsilon = 0}$. After this procedure, the error associated with the parameter $\epsilon$ is $\delta F /F <0.2\%$. Here, $\delta F$ is the difference between the values of $F_{\epsilon = 0}$ obtained by linear extrapolations over different sets of points. Typical values used in the calculations are $\epsilon = 10^{-4} \ldots 10^{-2}$.

With the $\epsilon$-regularization, the solutions can be calculated numerically until we reach $mr_h = m \beta_{\mathrm{BH}} / 4 \pi$, as shown by the red dots in Fig.~\ref{fig:mink-U}. Thus, we obtain the solution that describes the decay of the Unruh vacuum (Fig. \ref{fig:mink-U}b). This process also proceeds through the critical bubble for other values of $m\beta_{\mathrm{BH}}$. The solution at $\mathrm{Re}\, t_{\mathrm{A}} \rightarrow - \infty$ consists of right- and left-moving wave packets that correspond to the thermal and vacuum modes, respectively. As they approach $t_B$, these wave packets collide, and the field evolves nonlinearly, creating a critical bubble.

\begin{figure}[h]
	\centering
    \includegraphics[scale=1]{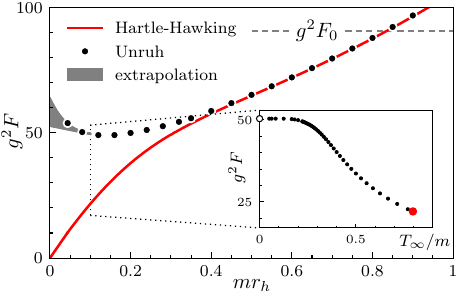}
	\caption{Main panel: Decay suppression as a function of the horizon radius $r_h$ for the Unruh vacuum (black dots) and the Hartle-Hawking vacuum (red line). Polynomial extrapolations in the limit $mr_h \rightarrow 0$ are all located in the gray area. The gray dashed line shows the decay suppression in flat space at zero temperature $F_0$. Inserted panel: black dots show decay suppression for various $T_{\infty}$ at fixed $T_{\mathrm{BH}} = 1/4\pi r_h$ and $mr_h = 0.1$. The red dot and black circle refer to Hartle-Hawking and Unruh vacua, respectively.}
	\label{fig:HH-U}
    \label{fig:test}
\end{figure}

The Unruh vacuum decay probability as a function of the horizon radius is shown by the black dots in Fig.~\ref{fig:HH-U}. There is a range of sizes, $mr_h = 0 \ldots 0.85$, in which BHs increase the decay probability as compared to the case of empty space, $g^2 F < g^2 F_0 \simeq 91$, but the suppression starts to increase in the limit of small $r_h$, and the probability remains exponentially suppressed for all BH sizes. The minimum suppression $g^2 F_{\mathrm{min}} \simeq 48$ is reached at $mr_h \simeq 0.12$. We used polynomial extrapolations of different orders in the limit $mr_h \rightarrow 0$, and all of their values sit within the gray area in Fig.~\ref{fig:HH-U}. The limiting values of the suppression are in the range $52<g^2 F <65$.  

The solutions and suppression values for the Unruh and Hartle-Hawking vacua also almost coincide for $mr_h \geq 0.4$. This happens because, as the BH size grows, the environment temperature in the Hartle-Hawking vacuum case becomes sufficiently low: $T_{\infty}/m = T_{\mathrm{BH}}/m \propto (mr_h)^{-1} \rightarrow 0$. This makes the Hartle-Hawking vacuum similar to the Unruh vacuum with $T_{\infty}/m = 0$. 

{\it 7.} In this letter, we have described a consistent approach for calculating the false-vacuum decay probability around a BH in (3+1)-dimensional spacetime. In the semiclassical approximation, the exponential suppression of the decay probability is estimated with solutions of classical equations on a specific complex time contour. We have described a procedure for obtaining numerical solutions of the problem for different values of the model parameters: the BH temperature, the thermal bath temperature, and the BH size. 
We verified that the solutions for Unruh and Hartle-Hawking vacua are continuously connected by varying the environmental temperature $T_{\infty}$ (Fig.~\ref{fig:test}), hence the found solutions sit on a branch that describes physical processes.

In this model, the vacuum decay proceeds through the critical bubble for both the Hartle-Hawking and Unruh vacua. In the range $mr_h \simeq 0\ldots 0.8$, BHs decrease the decay suppression down to $g^2 F_{\mathrm{min}} \simeq 48$; nevertheless, the decay probability remains exponentially suppressed in the limit of small horizon radii $r_h$. It is an effect of the four-dimensional model, since the energy density of the Hawking radiation drops with radial distance as $r^{-2}$, making it less probable for the field to create a true-vacuum bubble even at high temperature. Therefore, the assumption that BHs are sufficient catalysts of vacuum decay is most likely incorrect for other models with false vacuum potentials, including the SM Higgs sector.

{\it Acknowledgments.}  The work is supported in part by the scientific program of the National Center for Physics and Mathematics, section 5 "Particle Physics and Cosmology", stage 2026-2027.  R.G. thanks the Theoretical Physics and Mathematics Advancement Foundation “BASIS” for the student fellowship under the contract 24-2-10-40-1.


\addcontentsline{toc}{chapter}{\bibname}
\bibliographystyle{apsrev4-1}
\bibliography{references}


\end{document}